\documentclass[12pt,a4paper]{article}
\usepackage{times}
\usepackage{a4wide}
\usepackage{amsfonts}
\usepackage{amssymb}
\usepackage{amsmath}
\usepackage{bbm}
\usepackage{ifpdf}
\ifpdf
\usepackage[pdftex,unicode,implicit]{hyperref}
\hypersetup{%
  pdftitle    = {The supersymmetric black holes of $N=8$ supergravity},
  pdfkeywords = {gravity, quantum gravity, black hole, supersymmetry, N=8 supergravity},
  pdfauthor   = {Tom\'as Ort\'{\i}n and Carlos S. Shahbazi},
  plainpages  = true,
  colorlinks  = true,
  citecolor   = blue,
  urlcolor    = red,
  linkcolor   = black
}
\newcommand{\hepth}[1]{arXiv:{\tt
\href{http://www.arXiv.org/abs/hep-th/#1}{hep-th/#1}}}

\newcommand{\arxiv}[1]{{\tt
\href{http://www.arXiv.org/abs/#1}{arXiv:#1}}}
\else
  \usepackage[dvips]{graphicx}
  \usepackage[unicode,implicit]{hyperref}
  \newcommand{\hepth}[1]{arXiv:{\tt hep-th/#1}}

  \newcommand{\arxiv}[1]{{\tt arXiv:#1}}
\fi
\makeatletter
\@addtoreset{equation}{section}
\makeatother

\begin{document}

\begin{flushright}
\small
IFT-UAM/CSIC-12-51\\
June 13\textsuperscript{th}, 2012\\
\normalsize
\end{flushright}

\begin{center}

\vspace{1cm}

{\LARGE {\bf The supersymmetric black holes\\[.8cm] 
of $\mathcal{N}=8$ supergravity}}

\vspace{2.5cm}

\begin{center}

\renewcommand{\thefootnote}{\alph{footnote}}
{\sl\large Tom\'{a}s Ort\'{\i}n}
\footnote{E-mail: {\tt Tomas.Ortin [at] csic.es}}\, 
{\sl\large and C. S.~Shahbazi}
\footnote{E-mail: {\tt Carlos.Shabazi [at] uam.es}}
\renewcommand{\thefootnote}{\arabic{footnote}}

\vspace{1.5cm}

{\it Instituto de F\'{\i}sica Te\'orica UAM/CSIC\\
C/ Nicol\'as Cabrera, 13--15,  C.U.~Cantoblanco, 28049 Madrid, Spain}\\

\vspace{4cm}

\end{center}

{\bf Abstract}

\begin{quotation}

  {\small 
    Using the general results on the  classification of timelike
    supersymmetric solutions of all 4-dimensional $\mathcal{N}\geq 2$
    supergravity theories, we show how to construct all the supersymmetric
    (single- and multi-) black-hole solutions of $\mathcal{N}=8$ supergravity.
}

\end{quotation}

\end{center}

\setcounter{footnote}{0}

\newpage
\pagestyle{plain}



\section{Introduction}

For the last 20 years, black holes have been intensively studied in string
theory and supergravity with never-decreasing interest. A large part of effort
has been focused on two subjects: the construction of the most general
black-hole solutions of these theories and the understanding and computation
of different physical properties, specially the entropy, of the black-hole
solutions, following the seminal result of Strominger and Vafa
\cite{Strominger:1996sh}.

The attractor mechanism \cite{Ferrara:1995ih,Ferrara:1997tw} has provided a
bridge between these two subjects, allowing the computation of the entropy and
other black-hole properties on the black-hole horizon without the knowledge of
the complete black-hole solutions, at least in the extremal cases. In theories
with a very high degree of (super-) symmetry, though, it is not necessary to
use this mechanism and the entropy of the extremal black holes can be
determined requiring duality-invariance, correct dimensionality and
moduli-independence (which is a consequence of the attractor mechanism
\cite{Ferrara:1997tw}). In particular, the entropy of the extremal black holes
of $\mathcal{N}=8$ supergravity \cite{Cremmer:1979up,de Wit:1982ig} was found
in \cite{Kallosh:1996uy} to be given by the unique quartic invariant of the
$E_{7(7)}$ duality group. If we use the real basis 

\begin{equation}
\mathcal{Q} \equiv   
\left(
  \begin{array}{c}
  p^{ij} \\ q_{ij} \\  
  \end{array}
\right)\,  , 
\end{equation}

\noindent
for the charges, where the indices $i,j=1,\cdots,8$ transform homogeneously
under the $SL(8,\mathbb{R})\subset E_{7(7)}$ and each pair of indices is
antisymmetrized (so there are $28$ electric plus $28$ magnetic independent
charges), the quartic invariant is known as the Cartan invariant
$J_{4}(\mathcal{Q})$ \cite{CARTAN}

\begin{equation}
 \label{Cartan} 
J_{4}(\mathcal{Q})
= 
p^{ij}q_{jk}p^{kl}q_{li}
-\tfrac{1}{4} (p^{ij}q_{ij})^{2} 
+\tfrac{1}{96}\, \varepsilon_{ijklmnpq}p^{ij}p^{kl}p^{mn}p^{pq} 
+\tfrac{1}{96}\, \varepsilon^{ijklmnpq}q_{ij}q_{kl}q_{mn}q_{pq}\, . 
\end{equation}

In the complex basis, the quartic invariant is known as the Julia-Cremmer
invariant $\diamondsuit (\mathcal{Q})$ \cite{Cremmer:1979up}. They are
equal up to a sign \cite{Balasubramanian:1997az,Gunaydin:2000xr} and we will
not be concerned with its explicit form.

Although it has not been proven directly\footnote{To the best of our
  knowledge, not even within the FGK formalism of \cite{Ferrara:1997tw}.}, the
entropy formula for the extremal black holes of $\mathcal{N}=8$
supergravity

\begin{equation}
\label{eq:entropyformula}
S = \pi \sqrt{|J_{4}(\mathcal{Q})|}\, ,   
\end{equation}

\noindent
has passed all checks and, in particular, it has been shown to reproduce the
entropies of black holes of supergravity theories with $\mathcal{N}<8$
(specially $\mathcal{N}=2$) obtained by truncation of $\mathcal{N}=8$. For
supersymmetric black holes $J_{4}(\mathcal{Q})>0$ and one does not need to
take the absolute value.

One of the main obstructions for proving this formula is our lack of knowledge
of the general extremal black-hole solutions of $\mathcal{N}=8$ supergravity
as opposite to our complete knowledge of those of the $\mathcal{N}=2$ theories
\cite{Behrndt:1997ny,LopesCardoso:2000qm,Denef:2000nb,Bates:2003vx,Meessen:2006tu}. This,
and the standard lore that all the $1/8$ supersymmetric (the ones with a
potentially regular horizon) black-hole solutions of $\mathcal{N}=8$ are
supersymmetric black-hole solutions of some of the $\mathcal{N}=2$ truncations
of that theory (which seems to have been disproven by the explicit examples of
\cite{Bena:2011pi,Bena:2012ub}) explains why most of the literature on
$\mathcal{N}=8$ black holes deals with such truncations.

The supersymmetric black-hole solutions of $\mathcal{N}=2$ supergravity were
re-discovered in \cite{Meessen:2006tu} among the time-like supersymmetric
solutions of the theory, which were found by exploiting the integrability
conditions of the Killing spinor equations following Tod \cite{Tod:1983pm}
along the lines of \cite{Gauntlett:2002nw}. The same procedure was followed in
\cite{Meessen:2010fh} for all $\mathcal{N}\geq 2, d=4$ ungauged
supergravities, using the (almost) $\mathcal{N}$-independent formalism of
\cite{Andrianopoli:1996ve}, but the result, which we are going to explain in
the next section, looked too complicated to be used in the explicit
construction of the solutions, in spite to its similarity to the result found
in the $\mathcal{N}=2$ case.

We have recently realized, though, that the results found in
\cite{Meessen:2010fh} do permit the explicit construction of the metric of the
most general single and multi-black-hole solutions of ungauged $\mathcal{N}=8$
supergravity. The complications are restricted to the explicit construction of
the scalar fields. Thus, we are going to show how to construct the metrics of
the most general black holes ungauged $\mathcal{N}=8$ supergravity, but we
will not be able to provide a simple algorithm to find the scalar fields
corresponding to those solution. Nevertheless, the consistency of the formalism
ensures their existence and there is much that can be learned from the
metrics.

In the next section we are going to summerize the recipe found in
\cite{Meessen:2010fh} to construct all the timelike supersymmetric solutions
of $\mathcal{N}=8$ supergravity and in the next one we are going to give the
general form of the metric of these solutions, after which we will discuss the
black-hole case, showing how the entropy formula (\ref{eq:entropyformula})
arises for supersymmetric black holes and which of $E_{7(7)}$ invariants
studied in \cite{Andrianopoli:2011gy} actually arise in the two-center case.


\section{The timelike supersymmetric solutions of $\mathcal{N}=8$
  supergravity}

According to the results of \cite{Meessen:2010fh}, in order to construct a
timelike black-hole-type supersymmetric solution of $\mathcal{N}=8$
supergravity we may proceed as follows\footnote{We have included in this
  recipe, to simplify it, the vanishing of the ``hyperscalars''.}:

\begin{enumerate}
\item Choose an $x$-dependent rank-2, $8 \times 8$ complex antisymmetric
  $M_{IJ}$, These matrices must satisfy a number of constraints that are
  difficult to solve. This implies that, in practice, we cannot construct the
  most general matrices that satisfy them. Nevertheless, with those matrices
  we can proceed to the next step.

\item The scalars are encoded into the $56$-dimensional symplectic vector

\begin{equation}
(\mathcal{V}^{M}{}_{IJ})
=
\left( 
  \begin{array}{c}
f^{ij}{}_{IJ} \\ h_{ij\, IJ} \\
  \end{array}
\right)\, ,
\end{equation}

\noindent
antisymmetric in the \textit{local} $SU(8)$ indices $I,J=1,\cdots 8$. It
transforms in the fundamental (\textbf{56}) of $E_{7(7)}$ ($ij$ indices) and
as antisymmetric $U(8)$ tensor ($IJ$ indices), It satisfies\footnote{The
  symplectic product of two vectors $\langle \mathcal{A}\mid\mathcal{B}\rangle$
is defined by 
\begin{equation}
\langle \mathcal{A}\mid\mathcal{B}\rangle 
\equiv 
\mathcal{A}_{M}\mathcal{B}^{M}
\equiv
\mathcal{A}^{N}\mathcal{B}^{M}\Omega_{MN}\, ,
\end{equation}
where
\begin{equation}
(\Omega_{MN})
\equiv
\left(
  \begin{array}{cc}
 0 & \mathbbm{1}_{28\times 28} \\
-\mathbbm{1}_{28\times 28}   & 0 \\
  \end{array}
\right)\, ,
\end{equation}
is the skew metric of Sp$(56,\mathbb{R})$ that we use to lower (as above) or
raise symplectic indices.  
}

\begin{equation}
\langle \mathcal{V}_{IJ}\mid\mathcal{V}^{*\, KL}\rangle 
=
\tfrac{1}{2}f^{*\, ij\, KL}h_{ij\, IJ} 
- 
\tfrac{1}{2} h^{*}{}_{ij}{}^{KL}f^{ij}{}_{IJ}
=    
-2i\delta^{KL}{}_{IJ}\, ,
\hspace{1cm}
\langle \mathcal{V}_{IJ}\mid\mathcal{V}_{KL}\rangle  
= 
0\, ,
\end{equation}

Using the matrix $M_{IJ}$ chosen in the previous step, we define the real
symplectic vectors $\mathcal{R}^{M}$ and $\mathcal{I}^{M}$

\begin{equation}
\mathcal{R}^{M}+i\mathcal{I}^{M} 
\equiv  
\mathcal{V}^{M}{}_{IJ}\frac{M^{IJ}}{|M|^{2}}\, ,
\hspace{1cm}
|M|^{2} = M_{IJ}M^{IJ}\, .
\end{equation}

\noindent
These two are, by definition, $U(8)$ singlets (no $U(8)$ gauge-fixing
necessary) and only transform in the fundamental of $E_{7(7)}$.

\item The components of $\mathcal{I}$ are $56$ real functions
  $\mathcal{H}^{M}$ harmonic in the Euclidean $\mathbb{R}^{3}$ transverse
  space.

\item $\mathcal{R}$ is to be be found from $\mathcal{I}$ exploiting the
  redundancy in the description of the scalars by the sections
  $\mathcal{V}^{M}{}_{IJ}$\footnote{$\mathcal{V}^{M}{}_{IJ}$ uses $56^{2}$
    complex components to describe just $70$ physical scalars. The constraints
    that it satisfies imply a large number of relations between the
    components. The same is true for the components projected with
    $M_{IJ}$. This step is equivalent to the resolution of the
    \textit{stabilization equations} in $\mathcal{N}=2$ theories.}. Even with
  the knowledge of $M_{IJ}$ this is a very difficult step.

\item The metric is

\begin{equation}
\label{eq:themetric}
ds^{2} \; =\; e^{2U} (dt+ \omega)^{2} -e^{-2U}d\vec{x}^{\, 2}\, ,
\end{equation}

\noindent
where
  
\begin{eqnarray}
e^{-2U} 
& = &    
|M|^{-2}
=
\langle\,\mathcal{R}\mid \mathcal{I}\, \rangle
= 
\tfrac{1}{2}\mathcal{I}^{ij}\mathcal{R}_{ij}
-
\tfrac{1}{2}\mathcal{I}_{ij}\mathcal{R}^{ij}\, ,
\\
& & \nonumber \\
(d \omega)_{mn} 
& = &  
2\epsilon_{mnp}
\langle\,\mathcal{I}\mid \partial_{p}\mathcal{I}\, \rangle\, ,
\label{eq:omegaequation}
\end{eqnarray}

\noindent
and can be constructed automatically provided one has been given the harmonic
functions corresponding to $\mathcal{I}$ and $\mathcal{R}(\mathcal{I})$, quite
independently of the construction of these objects from $M_{IJ}$ and
$\mathcal{V}^{M}{}_{IJ}$. The same is true for the vector field strengths.

\item The vector field strengths are given by 

\begin{equation}
\mathcal{F}
=  
-{\textstyle\frac{1}{2}} d (\mathcal{R}\hat{V})   
-{\textstyle\frac{1}{2}}\star(\hat{V}\wedge d\mathcal{I}) \, ,  
\hspace{1.5cm}
\hat{V} = \sqrt{2} e^{2U}(dt+ \omega)\, . 
\end{equation}

\item The Vielbeins describing the scalars in the coset $E_{7(7)}/SU(8)$
  $P_{IJKL, \mu}$ are split into two complementary sets:

  \begin{equation}
    P_{IJKL}\, 
    \mathcal{J}^{I}{}_{[M}\mathcal{J}^{J}{}_{N} \mathcal{J}{}^{K}{}_{P} 
    \tilde{\mathcal{J}}{}^{L}{}_{Q]}\, , \,\,\,\,\,
    \mathrm{and}\,\,\,\,\,\, 
    P_{IJKL}\, \mathcal{J}^{I}{}_{[M} \tilde{\mathcal{J}}{}^{J}{}_{N}
    \tilde{\mathcal{J}}{}^{K}{}_{P} \tilde{\mathcal{J}}{}^{L}{}_{Q]}\, ,
  \end{equation}

\noindent
where we have defined the projectors 

\begin{equation}
\mathcal{J}^{I}{}_{J} \equiv \frac{2 M^{IK} M_{JK}}{|M|^{2}}\, ,
\hspace{1cm}
\mathcal{J}^{I}{}_{J}=\delta^{I}{}_{J}-\tilde{\mathcal{J}}^{I}{}_{J}\, .    
\end{equation}

  All those in the second set have been assumed to vanish from the start,
  since they would lead to a non-trivial metric in the transverse
  3-dimensional space, while those in the first set can in principle be found
  from $\mathcal{R}$ and $\mathcal{I}$, using the definitions of these vectors
  and of the Vielbein and the explicit form of chosen $M_{IJ}$, setting
  $\mathcal{I}^{M}=H^{M}(x)$ and confronting the third step: the resolution of
  the stabilization equations.

\end{enumerate}


\section{The metrics of the supersymmetric black-hole solutions of
  $\mathcal{N}=8$ supergravity}

If we want to construct the most general black-hole solutions of
$\mathcal{N}=8$ supergravity, the recipe demands a parametrization of the
space of all the matrices $M_{IJ}(x)$ that satisfy all the technical
requirements, which is very difficult to find.

We have realized, however, that this is a problem that we only need to solve
explicitly if we want to construct explicitly the scalar fields.  If we are
only interested in constructing the metric (and perhaps the vector fields) all
we really need is to assume that the problem has been solved and the resulting
$M_{IJ}(x)$ has been used to define $\mathcal{R}$ and $\mathcal{I}$.

One may naively think that both the explicit form of $M_{IJ}(x)$ and the
explicit expression of the components $\mathcal{V}^{M}{}_{IJ}$ are needed to
set up the stabilization equations and to solve them, finding $\mathcal{R}$ as
a function of $\mathcal{I}$.  Fortunately, this problem can be reformulated as
follows: with a real vector in the $\mathbf{56}$ of $E_{7(7)}$, $\mathcal{I}$,
we want to construct another one in the same representation which is a
non-trivial function of the former, $\mathcal{R}(\mathcal{I})$. For a single
$\mathcal{I}$, there is a unique way of constructing a $\mathbf{56}$ from
another $\mathbf{56}$, provided by the Jordan triple product\footnote{The
  Jordan triple product of three different $\mathbf{56}$s is defined only up
  to terms proportional to the symplectic products of two of the three
  $\mathbf{56}$s. The ambiguity disappears when we consider them to be equal,
  since the symplectic products will automatically vanish.}. Thus,
$\mathcal{R}^{M}(\mathcal{I})$ must be given by

\begin{equation}
  \mathcal{R}^{M}(\mathcal{I}) \sim (\mathcal{I},\mathcal{I},\mathcal{I})^{M}\, ,  
\end{equation}

\noindent
where

\begin{equation}
  \begin{array}{rcl}
    (\mathcal{I},\mathcal{I},\mathcal{I})^{ij}
    & = &  
    \tfrac{1}{2}
    \mathcal{I}^{ik}\mathcal{I}_{kl}\mathcal{I}^{lj}
    +\tfrac{1}{8}
    \mathcal{I}^{ij}\mathcal{I}_{kl}\mathcal{I}^{kl}
    -\tfrac{1}{96}\varepsilon^{ijklmnpq}\mathcal{I}_{kl}\mathcal{I}_{mn}\mathcal{I}_{pq}\, ,
    \\    
    & & \\
    (\mathcal{I},\mathcal{I},\mathcal{I})_{ij}
    & = &  
    -\tfrac{1}{2}
    \mathcal{I}_{ik}\mathcal{I}^{kl}\mathcal{I}_{lj}
    -\tfrac{1}{8}
    \mathcal{I}_{ij}\mathcal{I}_{kl}\mathcal{I}^{kl}
    +\tfrac{1}{96}\varepsilon_{ijklmnpq}\mathcal{I}^{kl}\mathcal{I}^{mn}\mathcal{I}^{pq}\, .
    \\    
  \end{array}
\end{equation}

To determine the proportionality factor we must first take into account that
we expect the $\mathcal{R}^{M}(\mathcal{I})$ to be homogenous of first order
in $\mathcal{I}$, which requires that we divide
$(\mathcal{I},\mathcal{I},\mathcal{I})$ by an $E_{7(7)}$ invariant (to
preserve the symmetry properties) homogenous of second degree in
$\mathcal{I}$, which can only be $\sqrt{J_{4}(\mathcal{I})}$.

We, thus, conclude that, up to a normalization constant $\beta$ to be
determined later, the solution to the stabilization equations of
$\mathcal{N}=8$ supergravity defined in the previous section is

\begin{equation}
\label{eq:R}
  \mathcal{R}^{M}(\mathcal{I}) = 
  \beta \frac{(\mathcal{I},\mathcal{I},\mathcal{I})^{M}}{\sqrt{J_{4}(\mathcal{I})}}\, ,  
\end{equation}

\noindent
which is our main result and allows the complete construction of the metrics
of all the supersymmetric black holes of the theory. 

Actually, since, as we are going to show in the next section, $\beta=2$,
$\mathcal{R}^{M}$ coincides exactly with the \textit{Freudenthal
  dual}\footnote{We thank M.~Duff and L.~Borsten for pointing out this fact to
  us.} of $\mathcal{I}^{M}$, which we can denote by $\tilde{I}^{M}$ defined in
\cite{Borsten:2009zy}. The Freudenthal dual $\tilde{Q}$ enjoys several
remarkable properties. Firstly,

\begin{equation}
\label{eq:firstproperty}
\langle\, \tilde{\mathcal{Q}} \mid \mathcal{Q}\, \rangle  = 2
J_{4}(\mathcal{Q})\, ,   
\end{equation}

\noindent
which follows from the property of the Jordan triple product

\begin{equation}
  \langle\, (\mathcal{Q},\mathcal{Q},\mathcal{Q})\mid \mathcal{Q}\, \rangle  
  =
  J_{4}(\mathcal{Q})\, .
\end{equation}

Secondly, 

\begin{equation}
\label{eq:secondproperty}
\tilde{\tilde{\cal Q}} = -\mathcal{Q}\, ,  
\end{equation}

\noindent
which eliminates a possible solution to the stabilization equations (namely
$\mathcal{R}^{M}=\tilde{\tilde{\cal I}}^{M}$) because $e^{-2U} = \langle\
\mathcal{R} \mid\mathcal{I} \rangle$ would vanish identically.

Thirdly,

\begin{equation}
\label{eq:thirdproperty}
J_{4}(\tilde{\cal Q}) =  J_{4}(\mathcal{Q})\, . 
\end{equation}

Finally, in \cite{Ferrara:2011gv} (where the definition of Freudenthal dual
was generalized to all $\mathcal{N}\geq 2$ theories) is has been shown to be a
symmetry of the space of critical points of the black-hole potential
introduced in \cite{Ferrara:1997tw}.

Thus, following the recipe, and choosing some harmonic functions $H^{M}(x)$,
the metric function $e^{-2U}$ is always given by 

\begin{equation}
\label{eq:metricfunction}
e^{-2U} =  \beta \sqrt{J_{4}(H)}\, ,  
\end{equation}

\noindent
and the 1-form $\omega$ is always given by the solution to

\begin{equation}
\label{eq:omega}
(d \omega)_{mn} 
 =   
\varepsilon_{mnp}
\left(\mathcal{I}_{ij}\partial_{p}\mathcal{I}^{ij}
-
\mathcal{I}^{ij}\partial_{p}\mathcal{I}_{ij}\right)\, .
\end{equation}

\noindent
the vector field strengths follow from the general formula and the scalars, as
mentioned before, cannot be easily recover, even if we now introduce an
$M_{IJ}$ with all the required properties. This is an evident shortcoming of
this procedure, but we believe it is compensated by the possibility of
studying explicitly the general black-hole metric.

Observe that, as expected, $\mathcal{R}_{M}$ can be obtained from the metric
function as

\begin{equation}
2\mathcal{R}_{M}(\mathcal{I}) = \frac{\partial\, e^{-2U}}{\partial
  \mathcal{I}^{M}}\, .  
\end{equation}

Furthermore, observe that the expression that we have given for the metric
function reduces to those found in \cite{Ferrara:2006yb} for all the magic
$\mathcal{N}=2$ truncations of $\mathcal{N}=8$ supergravity and another simple
truncation also reduces it to that of the well-known $STU$ model. The solution
to the stabilization equations of the 4-dimensional supergravities with
duality groups of type $E7$
\cite{Ferrara:2011dz,Ferrara:2011aa,Ferrara:2012qp} is also given by an
analogous expression.

In the next sections we analyze what these formulae mean for 1- and
2-center solutions.


\section{Single supersymmetric black-hole solutions}

To study more closely these black-hole metrics it is convenient to introduce
the so-called $\mathbb{K}$-tensor \cite{Marrani:2010de,Andrianopoli:2011gy},
which is associated to the completely symmetric linearization of the Cartan
invariant performed in \cite{Faulkner} (see \cite{Kallosh:2012yy} for more
details):

\begin{equation}
\label{eq:J4prime}
  \begin{array}{rcl}
J^{\prime}_{4}(\mathcal{Q}_{1},\mathcal{Q}_{2},\mathcal{Q}_{3},\mathcal{Q}_{4})
& \equiv &
\tfrac{1}{6}
\mathrm{Tr}_{SL(8,\mathbb{R})}
\left\{
 p_{1} \cdot  q_{2} \cdot p_{3} \cdot q_{4} 
+p_{1} \cdot q_{3} \cdot p_{4} \cdot q_{2}
+p_{1} \cdot q_{4} \cdot p_{2} \cdot q_{3}
+(p\leftrightarrow q)
\right\}
\\
& & \\
& &   
-\tfrac{1}{12}
\left\{
[\mathcal{Q}_{1}\mid \mathcal{Q}_{2} ] [\mathcal{Q}_{3} \mid \mathcal{Q}_{4} ]
+
[\mathcal{Q}_{1}\mid \mathcal{Q}_{3} ] [\mathcal{Q}_{2} \mid \mathcal{Q}_{4} ]
+
[\mathcal{Q}_{1}\mid \mathcal{Q}_{4} ] [\mathcal{Q}_{2} \mid \mathcal{Q}_{3} ]
\right\}
\\
& & \\
& & 
+\tfrac{1}{4}
\left[
\mathrm{Pf}_{SL(8,\mathbb{R})}||p_{1} p_{2} p_{3} p_{4}||
+
(p\leftrightarrow q)
\right]\, ,
\end{array}
\end{equation}

\noindent
where $\mathrm{Tr}_{SL(8,\mathbb{R})}$ stands for the trace of the products of
$p$ and $q$ matrices (always one upper and one lower index), we have defined,
for convenience, the symmetric product

\begin{equation}
[\mathcal{Q}_{1}\mid \mathcal{Q}_{2} ]
\equiv  
-\tfrac{1}{2}\mathrm{Tr}_{SL(8,\mathbb{R})}[p_{1}\cdot q_{2} +
(p\leftrightarrow q) ]\, ,  
\end{equation}

\noindent
and

\begin{equation}
\label{Pf}
  \begin{array}{rcl}
\mathrm{Pf}||p_{1} p_{2} p_{3} p_{4}||
& \equiv & 
\tfrac{1}{4!}
\varepsilon_{ijklmnop}
p_{1}^{ij}p_{2}^{kl}p_{3}^{mn}p_{4}^{op}\, ,
\\
& & \\
\mathrm{Pf}||q_{1} q_{2} q_{3} q_{4}||
& \equiv & 
\tfrac{1}{4!}
\varepsilon^{ijklmnop}
q_{1\, ij}q_{2\, kl}q_{3\, mn}q_{4\, op}\, .
\end{array}
\end{equation}

The $\mathcal{K}$-tensor can be defined by its contraction with four different
fundamentals:
\begin{equation}
\label{us}
 \mathbb{K}_{MNPQ} \mathcal{Q}_{1}{}^{M} \mathcal{Q}_{2}{}^{N}
 \mathcal{Q}_{3}{}^{P} \mathcal{Q}_{4}{}^{Q}
\equiv
J^{\prime}_{4}(\mathcal{Q}_{1},\mathcal{Q}_{2},\mathcal{Q}_{3},\mathcal{Q}_{4})
\, ,
\end{equation}

\noindent
and, since $J^{\prime}_{4}$ is completely symmetric in the four
$\mathbf{56}$s, the $\mathbb{K}$-tensor is also completely symmetric in the
four symplectic indices 

\begin{equation}
  \mathbb{K}_{MNPQ}= \mathbb{K}_{(MNPQ)}\, .
\end{equation}

By construction

\begin{equation}
J^{\prime}_{4}(\mathcal{Q},\mathcal{Q},\mathcal{Q},\mathcal{Q})
=
J_{4}(\mathcal{Q})
=  
\mathbb{K}_{MNPQ} \mathcal{Q}^{M} \mathcal{Q}^{N}
 \mathcal{Q}^{P} \mathcal{Q}^{Q}\, ,
\end{equation}

\noindent
and the Jordan triple product can be also written in terms of this tensor as

\begin{equation}
(\mathcal{Q},\mathcal{Q},\mathcal{Q})^{M} =   
\mathbb{K}^{M}{}_{NPQ} \mathcal{Q}^{N}
 \mathcal{Q}^{P} \mathcal{Q}^{Q}\, ,
\end{equation}

\noindent
so we can write the symplectic vector $\mathcal{R}$ (\ref{eq:R}) and the
metric function $e^{-2U}$ (\ref{eq:metricfunction}) in the more useful form

\begin{eqnarray}
 \mathcal{R}_{M} 
& = &
{\displaystyle\beta\frac{\mathbb{K}_{MNPQ}  H^{N} H^{P} H^{Q}}{\sqrt{J_{4}(H)}}}\, ,
\\
& & \nonumber \\
e^{-2U} 
& = &
\beta \sqrt{\mathbb{K}_{MNPQ} H^{M} H^{N} H^{P} H^{Q}}\, .
\label{eq:metricfunction}
\end{eqnarray}

Single, extremal, static ($\omega=0$) black-hole solutions are associated to
harmonic functions of the form

\begin{equation}
H^{M}= A^{M} + \frac{\mathcal{Q}^{M}/\sqrt{2}}{r}\, ,  
\hspace{1cm}
r \equiv |\vec{x}|\, ,
\end{equation}

\noindent
where the $A^{M}$ are constants to be determined in terms of the
physical constants of the solution. This is done by requiring asymptotic
flatness and absence of sources of NUT charge and using the relation between
these constants and the asymptotic values of the scalars (which we do not know
explicitly). This means that we will not be able to find the general form of
these constants. Nevertheless, let us see how far we can go.

Asymptotic flatness implies

\begin{equation}
|M_{\infty}|^{-2} 
=
\langle\, \mathcal{R}_{\infty} \mid \mathcal{I}_{\infty} \,\rangle
=
e^{-2U_{\infty}}
=
1\, ,
\end{equation}

\noindent
and requires the normalization

\begin{equation}
\label{eq:KA4=1}
\mathbb{K}_{MNPQ} A^{M} A^{N} A^{P} A^{Q}=
\beta^{-2}\, .
\end{equation}
 
The absence of sources of NUT charge follows from setting $\omega=0$ in
Eq.~(\ref{eq:omegaequation}):

\begin{equation}
\label{eq:AQ=0}
0
=
\langle\, A \mid \mathcal{Q} \, \rangle
= 
\Im{\rm m}\, \left(\mathcal{Z}_{\infty\, IJ}M^{IJ}_{\infty} \right)\, ,  
\end{equation}

\noindent
where we have used the definition of $\mathcal{I}$, we have also used
asymptotic flatness and the definition of the central charge matrix of
$\mathcal{N}=8$ supergravity

\begin{equation}
\mathcal{Z}_{IJ}\equiv \langle\, \mathcal{V}_{IJ} \mid \mathcal{Q} \,
\rangle\, .  
\end{equation}

The projection

\begin{equation}
\mathcal{Z} \equiv \tfrac{1}{\sqrt{2}}\mathcal{Z}_{IJ}\frac{M^{IJ}}{|M|^{2}}\, ,
\end{equation}

\noindent
plays the r\^ole of central charge for the solutions associated to $M^{IJ}$,
which projects in the $U(8)$ directions in which supersymmetry is preserved.
As shown in \cite{Ortin:2011vm}, it drives the flow of the metric function
(but not that of the $\mathcal{N}=8$ scalars).  The condition of vanishing NUT
charge can be written in the form

\begin{equation}
N = \Im{\rm m}\, \mathcal{Z}_{\infty}=0\, ,   
\end{equation}

\noindent
as in an $N=2$ theory with central charge $\mathcal{Z}$. As we are going to
see the mass of the black hole is given by the real part of
$\mathcal{Z}_{\infty}$ which coincides with the absolute value (because the
imaginary part vanishes)\footnote{Entirely analogous expressions have been
  given in \cite{Ferrara:2006yb} for the masses of the black holes of the
  magic $\mathcal{N}=2$ truncations of $\mathcal{N}=8$ supergravity.}:

\begin{equation}
M = 
| \mathcal{Z}_{\infty}|
=
\Re{\rm e}\, \mathcal{Z}_{\infty} 
=   
\tfrac{1}{\sqrt{2}}\langle\, \mathcal{R}_{\infty} \mid \mathcal{Q} \, \rangle
= 
\tfrac{1}{\sqrt{2}}\beta^{2}
\mathbb{K}_{MNPQ} A^{M} A^{N} A^{P}
\mathcal{Q}^{Q}\, .
\end{equation}

Taking these conditions and relations into account\footnote{We will have to
  impose additional conditions, like the positivity of the mass, to ensure the
  regularity of the metric.}, we find that the metric function has the form

\begin{equation}
e^{-2U}
=
\sqrt{
1+\frac{4M}{r} 
+
\frac{3\beta^{2}
\mathbb{K}_{MNPQ}A^{M}A^{N}\mathcal{Q}^{P}\mathcal{Q}^{Q}}{r^{2}}
+
\frac{\sqrt{2}\beta^{2} 
\mathbb{K}_{MNPQ}A^{M}\mathcal{Q}^{N}\mathcal{Q}^{P}\mathcal{Q}^{Q}}{r^{3}}
+\frac{\beta^{2} J_{4}(\mathcal{Q})/4}{r^{4}} 
}\, .  
\end{equation}

The asymptotic behavior confirms the identification of the mass parameter,
which, as all the other coefficients of the $1/r^{n}$ terms in the square root
(in particular $J_{4}(\mathcal{Q})$), has to be positive for the metric to be
regular. In the near-horizon limit $r\rightarrow 0$, the last term dominates
the metric function and we recover the well-known entropy formula
(\ref{eq:entropyformula}) setting $\beta=2$. The coefficients of $1/r^{2}$ and
$1/r^{3}$ do not have a simple expression in terms of the physical parameters.


\section{Supersymmetric 2-center solutions}

Multicenter solutions  can be constructed by choosing harmonic functions with
several poles, as in $\mathcal{N}=2$ theories \cite{Denef:2000nb,Bates:2003vx},  

\begin{equation}
H^{M}= A^{M} +\sum_{a}
\frac{\mathcal{Q}^{M}_{a}/\sqrt{2}}{|\vec{x}-\vec{x_{a}}|}\, ,  
\end{equation}

\noindent
and tuning the parameters $A^{M},\mathcal{Q}^{M}_{a},\vec{x}_{a}$,so the
integrability conditions of the equation for $\omega$ (\ref{eq:omegaequation})

\begin{equation}
  \langle\,  A \mid \mathcal{Q}_{a}\, \rangle
  +\sum_{b} \frac{\langle\,  \mathcal{Q}_{b} \mid \mathcal{Q}_{a}\,
    \rangle/\sqrt{2}}{|\vec{x}_{a}-\vec{x}_{b}|}  =0\, .
\end{equation}

\noindent
Summing the above equations over $a$ and taking into account the antisymmetry
of the symplectic product, we find that the constants $A$, apart from
satisfying (\ref{eq:KA4=1}), also satisfy the condition (\ref{eq:AQ=0}) where
$\mathcal{Q}=\sum_{a}\mathcal{Q}_{a}$. 

When these equations are satisfied, $\omega$ exists and describes the total
angular momentum of the multi-black-hole system, just as in the
$\mathcal{N}=2$ cases, since the equations are identical.

The (square of) the metric function will contain many terms, up to order
$|\vec{x}-\vec{x}_{a}|^{-4}$.  The term of order $|\vec{x}-\vec{x}_{a}|^{-1}$
has the coefficient

\begin{equation}
M_{a}
\equiv
2\sqrt{2}
\mathbb{K}_{MNPQ} A^{M} A^{N} A^{P}
\mathcal{Q}_{a}^{Q}\, ,
\end{equation}

\noindent
which corresponds to the mass that the $a^{\rm th}$ center if it was
isolated. The mass of the solution is the sum of these parameters
$M=\sum_{a}M_{a}$. 
 
The coefficient of $|\vec{x}-\vec{x}_{a}|^{-n}|\vec{x}-\vec{x}_{b}|^{-m}$ with
$m+n=4$ is one the five quartic invariants listed in
\cite{Andrianopoli:2011gy} for 2-center solutions
\begin{equation}
  \begin{array}{rcl}
I_{+2}
& = & 
 \mathbb{K}_{MNPQ} \mathcal{Q}_{a}{}^{M} \mathcal{Q}_{a}{}^{N}
 \mathcal{Q}_{a}{}^{P} \mathcal{Q}_{a}{}^{Q}
=
J^{\prime}_{4}(\mathcal{Q}_{a},\mathcal{Q}_{a},\mathcal{Q}_{a},\mathcal{Q}_{a})
=
J_{4}(\mathcal{Q}_{a})\, ,
\\
& & \\
I_{+1}
& = & 
 \mathbb{K}_{MNPQ} \mathcal{Q}_{a}{}^{M} \mathcal{Q}_{a}{}^{N}
 \mathcal{Q}_{a}{}^{P} \mathcal{Q}_{b}{}^{Q}
=
J^{\prime}_{4}(\mathcal{Q}_{a},\mathcal{Q}_{a},\mathcal{Q}_{a},\mathcal{Q}_{b})\, ,
\\
& & \\
I_{0}
& = & 
 \mathbb{K}_{MNPQ} \mathcal{Q}_{a}{}^{M} \mathcal{Q}_{a}{}^{N}
 \mathcal{Q}_{b}{}^{P} \mathcal{Q}_{b}{}^{Q}
=
J^{\prime}_{4}(\mathcal{Q}_{a},\mathcal{Q}_{a},\mathcal{Q}_{b},\mathcal{Q}_{b})\, ,
\\
& & \\
I_{-1}
& = & 
\mathbb{K}_{MNPQ} \mathcal{Q}_{a}{}^{M} \mathcal{Q}_{b}{}^{N}
 \mathcal{Q}_{b}{}^{P} \mathcal{Q}_{b}{}^{Q}
=
J^{\prime}_{4}(\mathcal{Q}_{a},\mathcal{Q}_{b},\mathcal{Q}_{b},\mathcal{Q}_{b})\, ,
\\
& & \\
I_{-2}
& = & 
\mathbb{K}_{MNPQ} \mathcal{Q}_{b}{}^{M} \mathcal{Q}_{b}{}^{N}
\mathcal{Q}_{b}{}^{P} \mathcal{Q}_{b}{}^{Q}
=
J^{\prime}_{4}(\mathcal{Q}_{b},\mathcal{Q}_{b},\mathcal{Q}_{b},\mathcal{Q}_{b})\, ,
=
J_{4}(\mathcal{Q}_{b})\, .
\end{array}
\end{equation}

\noindent
The $I_{+2}$ $I_{-2}$ give the contributions of each center to the entropy.

With more than two centers, other combinations will appear based on the
quartic invariant. The sextic invariant found in \cite{Andrianopoli:2011gy}
does not seem to occur in these solutions.

\section{Conclusions}

In this paper we have shown how to construct the most general supersymmetric
black holes of ungauged $\mathcal{N}=8$ supergravity. While it is true that
one could have guessed the form of the metric function, given by the Cartan
quartic invariant, the relation between the harmonic functions and the rest of
the fields and the equation for the 1-form $\omega$ would have to be derived
by solving very complicated equations of motion. This part of the job had
already been done in \cite{Meessen:2010fh} and, as discussed in the previous
sections, one only had to solve the \textit{stabilization equations} for the
theory. Using the knowledge available in the literature we have been able to
solve these equations in terms of the unique Jordan triple
product\footnote{Unique for a single $\mathbf{56}$.}, which has allowed us to
prove (rather than guess) the result $e^{-2U} =2\sqrt{J_{4}(H)}$.

As we have shown, the general results of \cite{Meessen:2010fh}, together with
the solutions of the stabilization equations allows us to construct any
supersymmetric multi-black-hole solution as well.

It is natural to ask how the extremal non-supersymmetric and non-extremal
black holes of the theory can be found.

In a recent paper \cite{Galli:2011fq} it has been argued that all the black
holes of a given theory should have the same form as functions of some
elementary building blocks that are harmonic functions in the extremal cases
(supersymmetric or not) and linear combinations of hyperbolic sines and
cosines in the non-extremal cases. These building blocks should transform
linearly under the duality group (preserving harmonicity or ``linear
hyperbolicity'').

In $\mathcal{N}=2$, $d=4,5$ supergravities, there are natural candidates for
these building blocks and the conjecture was successfully tested in several
examples in the above reference and \cite{Meessen:2011bd}. Actually, for these
theories, it can be shown that this is always the case
\cite{Mohaupt:2009iq,Mohaupt:2010fk,Mohaupt:2011aa,Meessen:2011aa,Meessen:2012su},
since there is always a change of variables from the conventional ones to the
building blocks which are harmonic in the extremal cases.

In $\mathcal{N}=8$ supergravity a proof of this kind is not available but we
can repeat the arguments of \cite{Galli:2011fq} to argue that, at least for
single, static black holes, the metric should always have the form
(\ref{eq:themetric}) with $\omega=0$ and $e^{-2U}$ given by
(\ref{eq:metricfunction}) with $\beta=2$ and with the $H^{M}$ given by radial
functions with different profiles. 

It is clear that more work is necessary to test this possibility which we
intend to explore in a forthcoming publication.


\section*{Acknowledgments}

TO would like to thank Patrick Meessen for very useful conversations his
collaboration over the years.  This work has been supported in part by the
Spanish Ministry of Science and Education grant FPA2009-07692, the Comunidad
de Madrid grant HEPHACOS S2009ESP-1473 and the Spanish Consolider-Ingenio 2010
program CPAN CSD2007-00042. The work of CSS has been supported by a CSIC
JAE-predoc grant JAEPre 2010 00613. TO also wishes to thank M.M.~Fern\'andez
for her permanent support.


\end{document}